\documentclass[aps,pra,onecolumn,superscriptaddress,10pt]{revtex4}

\usepackage{latexsym}
\usepackage{amsmath}
\usepackage{dcolumn}
\usepackage{url}

\usepackage{setspace}
\usepackage{natbib}

\usepackage{upgreek}

\usepackage{mathrsfs}
\usepackage{bbm}
\usepackage{stmaryrd}
\usepackage{bm}

\usepackage{amsfonts}
\usepackage{graphicx}        

\newcommand{\rme}{{\mathrm{e}}}

\newcommand{\rmpp}{{\mathrm{p}}}
\newcommand{\rmw}{{\mathrm{w}}}
\newcommand{\rmd}{{\mathrm{d}}}
\newcommand{\rmi}{{\mathrm{i}}}
\newcommand{\rmD}{{\mathrm{D}}}

\newcommand{\bfk}{\mathbf{k}}

\newcommand{\bfm}{\mathbf{m}}
\newcommand{\bfn}{\mathbf{n}}

\newcommand{\bfr}{\mathbf{r}}

\newcommand{\bfv}{\mathbf{v}}

\newcommand{\cMs}{{\mathcal{M}}_{\mathrm{smooth}}}

\begin{document}

\def\Marseille{UMR 7345 CNRS, Aix-Marseille Universit\'e, 
campus Saint-J\'er\^ome, case 321, \\ 
av.\ esc.\  Normandie-Niemen, FR-13397 Marseille cedex 20}

\title[$N$-body description of Debye shielding and Landau damping]{$N$-body description of Debye shielding and Landau damping}

\author{D~F Escande, F Doveil and Yves Elskens}


\affiliation{\Marseille}
\email{dominique.escande@univ-amu.fr, fabrice.doveil@univ-amu.fr, yves.elskens@univ-amu.fr}

\begin{abstract}
This paper brings further insight into the recently published $N$-body description of Debye shielding and Landau damping [Escande D~F, Elskens Y and Doveil F 2014 \textit{Plasma Phys. Control. Fusion} \textbf{57}  025017]. Its fundamental equation for the electrostatic potential is derived in a simpler and more rigorous way. Various physical consequences of the new approach are discussed, and this approach is compared with the seminal one by Pines and Bohm [Pines D and Bohm D 1952 \textit{Phys. Rev.} \textbf{85} 338--353].
\newline \newline
{\bf{PACS numbers}}~:  \newline
      {52.20.-j}  {Elementary processes in plasmas}  \newline 
      {52.35.Fp}  {Plasma : electrostatic waves and oscillations} \newline
      {45.50.-j}  {Dynamics and kinematics of a particle and a system of particles}   \newline 
      {05.20.Dd}  {Kinetic theory} \newline



\end{abstract}


\maketitle

\section{Introduction}
\label{secIntro}

In order to discuss microscopic plasma physics, most textbooks start with invoking the $N$-body description of plasmas,
but they deem it too difficult to tackle, and they rapidly switch to kinetic equations.
The kinetic approach has proved extremely powerful to compute a wealth of plasma effects,
but it often hides some aspects of their mechanical nature~;
for instance, after the Vlasovian derivation of Landau damping,
most textbooks use simple mechanical models to provide physical interpretations of this damping.

However, the $N$-body description of plasmas has proved for more than six decades to be very powerful
at describing analytically effects like Debye shielding, Landau damping, spontaneous emission of Langmuir waves,
and at revealing the corresponding mechanical behavior of both particles and waves.
This started in 1952 with a seminal paper by Pines and Bohm \cite{PiBo}.
From the nineties on, there was a revival of the $N$-body approach to describe analytically wave-particle interaction in plasmas
\cite{AEE,EEbook}.

Last year, reference \cite{DbCb} provided a direct and simultaneous derivation of Debye shielding and Landau damping by an $N$-body description of the plasma.
This derivation works directly with Newton's second law and with a version of the electrostatic potential linearized from ballistic orbits of the particles~;
it does not use probabilistic arguments or partial differential equations.
However, in retrospect, the formulation of this paper now appears to be unnecessarily intricate,
because it was still too close to the heuristic derivation of the theory~;
it also required an erratum \cite{DbCbCor}.

The heuristic derivation substituted the discrete summations over the particles, which are intrinsic to the $N$-body approach,
with integrals over a continuous velocity distribution, and was presented at the beginning of reference \cite{DbCb}.
In retrospect, this substitution turns out to be an efficient shortcut to the simultaneous derivation of Debye shielding and Landau damping, but this non rigorous step can be avoided by considering the initial $N$-body distribution as close to a set of monokinetic arrays of particles. This other approach was presented in a second part of reference \cite{DbCb}. In strong resonance with Dawson's seminal multi-beam fluid theory \cite{Dawson60}, this second approach shows that Landau damping is due to phase mixing \`a la van Kampen \cite{DbCb}, while the first approach leads to the Vlasovian expression introduced by Landau that hides this fact.

Sections \ref{FEP} and \ref{SCP} (and the appendix) provide a shorter and more rigorous version of the approach of reference \cite{DbCb} using a smooth distribution function.
The calculation reduces to a mere exercise using Fourier series and Laplace transform.
Section \ref{DSLD} discusses the various physical consequences of the present $N$-body approach. Section \ref{BGP} is devoted to the comparison of the present $N$-body approach with the 1952 one by Pines and Bohm \cite{PiBo}.
This latter work is often overlooked by plasma physics textbooks and was overlooked also by the present authors in reference \cite{DbCb}.
The main technical difference of the present theory with reference \cite{PiBo} is the use of the Laplace transform. It turns out to have the same heuristic power as in the Vlasovian approach, and enables a compact and straightforward simultaneous derivation of Debye shielding and Landau damping without any a priori physical intuition.

\section{Simpler derivation of the fundamental equation for the electrostatic potential}
\label{FEP}

As in reference \cite{DbCb}, we deal with the One Component Plasma (OCP) model \cite{Salp,Abe,BH}, which considers the plasma as infinite with spatial periodicity $L$ in three orthogonal directions with coordinates $(x,y,z)$, and made up of $N$ electrons in each elementary cube with volume $L^3$. Ions are present only as a uniform neutralizing background, enabling periodic boundary conditions. We now provide a shorter and more rigorous version of the derivation of reference \cite{DbCb} using a smooth distribution function. The actual calculation involves sums over the $N$ particles and over all Fourier components of the potential.
However, the principle of this calculation can be explained more simply by performing it formally for the unphysical case of a single electron acted upon by a single Fourier harmonic of its own field. This is done in the next subsection, while the complete derivation is in appendix.

\subsection{Single particle acted upon by a single Fourier harmonic of its own field}
\label{PSEPS}

While the Fourier decomposition of the periodic delta distribution corresponding to the charge density of an electron entails an infinity of harmonics, in this subsection we keep just a single one with wavevector $\bfk$ which provides through Poisson equation the single harmonic of the electrostatic potential
\begin{equation}
  \tilde{\varphi} (t)
  =
     -\frac{e}{\epsilon_0 k^2} \exp[- \rmi \bfk \cdot \bfr(t)],
\label{*phitildetotM}
\end{equation}
where $k = \|\bfk\|$, $\epsilon_0$ is the vacuum permittivity,
and $\bfr(t)$ is the position at time $t$ of the particle acting as a source with charge $- e$.
Let $\bfr_{0}$ and $\bfv$ be respectively the initial position and velocity of the particle, and let $\Delta \bm{r} (t)= \bfr (t) - \bfr_{0} - \bfv t$. We now compute a perturbative solution to the  dynamics.
To this end, setting $\bfr (t) = \bfr_{0} + \bfv t + \Delta \bm{r} (t)$ in equation (\ref{*phitildetotM}), we replace $\tilde{\varphi}$ with its expansion to first order in $\Delta \bm{r} (t)$
\begin{equation}
\tilde{\varphi}_{\rm lin} (t)
  =
  - \frac{ e}{\epsilon_0 k^2} \exp [- \rmi \bfk \cdot (\bfr_{0} + \bfv t)] \ [ 1 - \rmi \bfk \cdot \Delta \bm{r}(t)].
\label{PPhitildnj}
\end{equation}
We now use the time Laplace transform which maps a function $ g(t)$
to $\widehat{g}(\omega) = \int_0^{\infty}  g(t) \exp(\rmi \omega t) \rmd t$
(with $\omega$ complex). Since the arguments of functions are spelled explicitly,
from now on we omit diacritics for the Laplace (or Fourier) transformed quantities.
The time Laplace transform of equation (\ref{PPhitildnj}) is
\begin{equation}
\varphi_{\rm lin} (\omega)
  =
  - \frac{ e}{\epsilon_0 k^2} \exp [- \rmi \bfk \cdot \bfr_{0}] \ [ \frac{\rmi}
             {\omega -\bfk  \cdot \bfv} - \rmi \bfk \cdot \Delta \bm{r} (\omega - \bfk  \cdot \bfv) ],
\label{*phitildnj}
\end{equation}
where the Doppler shift $- \bfk \cdot \bfv$ comes from the linear dependence on $t$ of the exponent of equation\ (\ref{PPhitildnj}).

To compute $\Delta \bm{r} (\omega - \bfk  \cdot \bfv) $, we use Newton's equation, which writes
\begin{equation}
  \ddot{\bfr}
  = \frac{e}{m_\rme} \nabla \varphi_{\rm lin}(\bfr)
\label{*rsectot}
\end{equation}
for the particle with mass $m_\rme$.
The single harmonic of the electrostatic potential defined by equation\ (\ref{*phitildetotM}) implies
\begin{equation}
\varphi_{\rm lin}(\mathbf{r}) = \frac{1}{L^3} \ \tilde{\varphi}_{\rm lin} \ \exp(\rmi \bfk \cdot \mathbf{r}) ,
\label{*phiInv}
\end{equation}
where $\mathbf{r}$ now is the position of the particle as a massive object subjected to a force.

In order to derive our perturbative solution, we consider $k \| \Delta \bm{r} \|$ to be small in equation\ (\ref{PPhitildnj}),
so that the linearized particle dynamics defined by equation\ (\ref{*rsectot}) reduces to
\begin{equation}
  \Delta \ddot{\bm{r}}
  =
 \frac{\rmi e}{L^3 m_\rme} \ \bfk \
    \tilde{\varphi}_{\rm lin}(t) \exp[\rmi \bfk \cdot (\bfr_{0} + \bfv t)].
\label{*delrsec}
\end{equation}
Since $\Delta \bm{r} (0) =0$ and $\Delta \dot{\bm{r}} (0) =0$,
the time Laplace transform of equation\ (\ref{*delrsec}) is
\begin{equation}
  - \omega^2 \Delta \bm{r}(\omega)
  = \frac{\rmi e}{L^3 m_\rme} \
                \bfk \ \exp(\rmi \bfk \cdot \bfr_{0})
                  \ \varphi_{\rm lin}(\omega + \bfk \cdot \bfv),
\label{*rLapl}
\end{equation}
where the Doppler shift $\bfk \cdot \bfv$
comes from the linear dependence on $t$ in the exponent of equation (\ref{*delrsec}).
Computing $\Delta \bm{r} (\omega - \bfk  \cdot \bfv) $ in equation (\ref{*phitildnj}) from the expression of $\Delta \bm{r}(\omega)$ given by equation (\ref{*rLapl}) yields
\begin{equation}
\qquad
   k^2\varphi_{\rm lin}(\omega)
 - \frac{e^2}{m_\rme \epsilon_0 L^3} \,
 \bfk \cdot \bfk
  \ \frac{\varphi_{\rm lin}(\omega + \bfk \cdot \bfv - \bfk \cdot \bfv)}{(\omega - \bfk \cdot \bfv)^2}
                             \exp[\rmi (\bfk -\bfk) \cdot \bfr_{0}]
  =
  k^2 \varphi_{\rm lin}^{(\rm{bal})}(\omega) ,
\label{phihat}
\end{equation}
where
\begin{equation}
  \varphi_{\rm lin}^{(\rm{bal})}(\omega) =
      - \frac{\rmi e}{\epsilon_0 k^2} \frac{\exp[- \rmi \bfk  \cdot \bfr_{0}]}
             {\omega -\bfk  \cdot \bfv},
\label{phij0hat}
\end{equation}
is the time Laplace transform of the ballistic potential obtained by setting $\Delta \bm{r}(t)=0$ in equation (\ref{PPhitildnj}). We keep on purpose the terms $(\bfk -\bfk)\cdot \bfr_{0}$ and $\bfk \cdot \bfv - \bfk \cdot \bfv$ due to the two Doppler shifts,
in order to exhibit the structure of the equation when the full set of wavevectors is taken into account.

\subsection{Fundamental equation for the electrostatic potential}
\label{CEEP}

As derived in the appendix,
when the full set of wavevectors and particles is taken into account, equation (\ref{phihat}) becomes
\begin{equation}
   k_{\bfm}^2 \varphi_{\rm lin}(\bfm,\omega)
 - \frac{\omega_{\rmpp}^2}{N}
 \sum_{\bfn, \, k_{\bfn} b_{\mathrm{smooth}} \leq 1} \bfk_{\bfm} \cdot \bfk_{\bfn}
  \ \sum_{j = 1}^N \frac{\varphi_{\rm lin}(\bfn,\omega + \bfk_{\bfn} \cdot \bfv_{j} - \bfk_{\bfm} \cdot \bfv_{j})}{(\omega - \bfk_{\bfm} \cdot \bfv_{j})^2}
                             \exp[\rmi (\bfk_{\bfn}-\bfk_{\bfm}) \cdot \bfr_{j0}]
  =
  k_{\bfm}^2 \varphi_{\rm lin}^{(\rm{bal})}(\bfm,\omega) ,
\label{*phihat}
\end{equation}
where $\omega_{\rmpp} = [e^2 n / (m_\rme \epsilon_0)]^{1/2}$ is the plasma frequency
(with $n = N/L^3$),
where $\varphi_{\rm lin}(\bfm,\omega)$ is the time Laplace transform of the Fourier component of the electro\-static potential with wavevector $\bfk_{\bfm} = \frac{2 \pi}{L} \, \bfm$, and $\bfm = (m_x,m_y,m_z)$ is a vector with three integer components
(we comment below on the restriction $k_{\bfn} b_{\mathrm{smooth}} \leq 1$ making $\bfk_\bfn$ run over a bounded domain).
Note that $g(\bfm, \omega)^* = g(- \bfm, - \omega^*)$ for real $\bfm$ and complex $\omega$ if $g(\bfr, t)$ is real-valued.
We also defined
\begin{equation}
  \varphi_{\rm lin}^{(\rm{bal})}(\bfm,\omega) =
  \sum_{j = 1}^N \varphi_{{\rm{lin}},j}^{(\rm{bal})}(\bfm,\omega)
  ,
\label{*phi0hat}
\end{equation}
with
\begin{equation}
  \varphi_{{\rm{lin}},j}^{(\rm{bal})}(\bfm,\omega)
  = - \frac{\rmi e}{\epsilon_0 k_{\bfm}^2}
      \frac{\exp[- \rmi \bfk_{\bfm}  \cdot \bfr_{j0}]}
             {\omega -\bfk_{\bfm}  \cdot \bfv_j} ,
\label{*phij0hat}
\end{equation}
the ballistic potential related to particle $j$, i.e.\  the unshielded Coulomb potential of an electron with position $\bfr_j(t) = \bfr_{j0} + \bfv_j t$.

In this complete derivation, for any finite value of the $\| \Delta \bm{r} \|$'s due to the various particles, the linearization in the $\Delta \bm{r}$'s is justified for finite values of the wavenumber moduli $k_{\bfm} = \|\bfk_{\bfm}\|$ only.
Therefore, as in section 4.5 of reference \cite{DbCb}, we restrict the Fourier expansion of the Coulomb potential to $k_{\bfm}$'s such that $k_{\bfm} b_{\mathrm{smooth}} \leq 1$,
where $b_{\mathrm{smooth}} \ll \lambda_\rmD = [\epsilon_0 k_{\rm{B}} T / (n e^2)]^{1/2}$, the Debye length,
with $k_{\rm{B}}$ the Boltzmann constant, and $T$ the temperature~;
let us denote by $\cMs$ the set of such wavevectors.
Then, we may consider finite $\| \Delta \bm{r} \|$'s, with $\| \Delta \bm{r} \| \ll b_{\mathrm{smooth}}$.
The above truncation of the Fourier expansion smoothes the Coulomb potential. We notice that a similar smoothing of the $1/r$ singularity of this potential (or some penalization of very close approaches \cite{Kie14})
is performed in the mean-field derivation of the Vlasov equation \cite{NeunzertWick,Neunzert84, Dobru,BraunHepp,Spohn,ElsVla}.
In any case, the Vlasov equation cannot describe scales smaller than the inter-particle distance. Since the self-field due to the smoothed Coulomb potential vanishes, it is not necessary to exclude self-interaction in the complete derivation \cite{DbCbCor}.

Equation (\ref{*phihat}) is the fundamental equation of this paper.
\emph{This equation is of the type ${\mathcal{E}} \varphi_{\rm lin}=$ source term},
where ${\mathcal{E}}$ is a linear operator,
acting on the infinite dimensional array whose components are all the Doppler shifted $\varphi_{\rm lin}(\bfm,\omega)$'s.

\section{Smoothed linear operator of the potential equation }
\label{SCP}

We now approximate the granular (or empirical, Kolmogorov) distribution
$F(\bfr, \bfv) = \sum_{j = 1}^N \delta(\bfr - \bfr_{j0}) \  \delta(\bfv -  \bfv_{j})$
with a position and velocity distribution function $f(\bfr,\bfv)$
that is \emph{continuous} in $\bfr$, such that distribution $f-F$ yields a negligible contribution
when applied to space dependent functions which vary slowly on the scale $b_{\mathrm{smooth}} $.
This means that we use a version of $F$ coarse-grained on this scale.
On replacing the discrete sums over particles with integrals over the smooth distribution function $f(\textbf{r},\textbf{v})$ in its left hand side, equation (\ref{*phihat}) becomes
\begin{equation}
  k_{\bfm}^2 \varphi_{\rm cg}(\bfm,\omega)
  - \frac{\omega_{\rmpp}^2}{N} \sum_{\bfn}
        \textbf{k}_{\bfm} \cdot \textbf{k}_{\bfn}
    \int  \frac{\varphi_{\rm cg}(\bfn,
\omega + (\textbf{k}_{\bfn} - \textbf{k}_{\bfm}) \cdot \textbf{v})}
               {(\omega -\textbf{k}_{\bfm} \cdot \textbf{v})^2}
  \tilde{f}(\textbf{k}_{\bfn} - \textbf{k}_{\bfm},\textbf{v}) \ \rmd^3 \textbf{v}
= 
  k_{\bfm}^2 \varphi_{\rm lin}^{(\rm{bal})}(\bfm,\omega)
,
\label{phihatcg}
\end{equation}
where $\varphi_{\rm cg}$ is the smoothed version of $\varphi_{\rm lin}$, and $\tilde{f}$ is the spatial Fourier transform of $f$. We consider almost uniform initial spatial distributions of particles. So, $f$ is a spatially uniform distribution function $f_0(\textbf{v})$ ($\iint f_0(\bfv)\, \rmd^3 \bfr \, \rmd^3 \bfv = L^3 \int  f_0(\bfv)\, \rmd^3 \bfv = L^3 n = N$) plus a small perturbation $f_1(\textbf{r},\textbf{v})$ of the order of $\varphi_{\rm cg}$.
In linearizing equation (\ref{phihatcg}) for $\varphi_{\rm cg}$, $\tilde{f}(\textbf{k},\textbf{v})$ is substituted with $\tilde{f}_0(\textbf{k},\textbf{v})$ which vanishes for $\textbf{k} \neq {\mathbf{0}}$. Therefore, operator $\mathcal{E}$ becomes diagonal with respect to both $\bfm$ and $\omega$ (a complex quantity), and equation (\ref{phihatcg}) becomes
\begin{equation}
  \epsilon(\bfm,\omega) \, \varphi_{\rm cg}(\bfm,\omega)
  = \varphi_{\rm lin}^{(\rm{bal})}(\bfm,\omega),
\label{phihatL}
\end{equation}
with
\begin{equation}
  \epsilon(\bfm,\omega)
  = 1 - \frac{\omega_{\rmpp}^2 L^3}{N}
     \int \frac{f_0(\bfv) }{(\omega - \bfk_{\bfm}  \cdot \bfv)^2} \ \rmd^3 \bfv.
\label{eps}
\end{equation}

The validity of the above coarse-graining was discussed in section 4.5 of reference \cite{DbCb}. Here we only recall the main steps of the discussion. First, the presence of many particles in the Debye sphere justifies neglecting the non-diagonal terms in equation\ (\ref{*phihat}). Second, Dawson's work \cite{Dawson60} justifies the Vlasovian expression (\ref{eps}) for $\epsilon(\bfm,\omega)$ for a one-dimensional plasma. The generalization of Dawson's calculation to the three-dimensional case, though tedious, does not a priori involve any new conceptual difficulty. These results back up the above coarse-graining.

However, we now show these results come with conditions on the parameters of our problem. Invoking Dawson's calculation means that, when interested in a wave with phase velocity $\bfv_\rmw$ (with $\| \bfv_\rmw \| = v_\rmw$),
there are many particles with parallel velocities in the range $[v_\rmw - |\gamma_{{\rm L}}|/k, v_\rmw + |\gamma_{{\rm L}}|/k]$,
where $\gamma_{{\rm L}}$ is the Landau growth or damping rate of the wave, and $k$ the modulus of its wavenumber.
Indeed, this range is the one where the phase mixing \`a la van Kampen is occurring, and where the synchronization of particles with the wave brings the change of particle momentum inducing Landau damping or growth of this wave \cite{EZE,EEbook,EFields,Houches,DovEsMa}. If the initial distribution of particles is meant to approximate a smooth distribution $f_0(\textbf{v})$, this means that the length $L$ of the system is taken large enough to fulfill the condition ``many particles with parallel velocities in the range $[v_\rmw - |\gamma_{{\rm L}}| / k, v_\rmw + |\gamma_{{\rm L}}| / k]$".
However, whatever $L$, for $v$ large enough, one should not use $\epsilon(\bfm,\omega)$, but its discrete counterpart computed with the $N$-body distribution function.

\section{Debye shielding, Langmuir waves and Landau damping}
\label{DSLD}

To keep this paper self-contained, we now discuss
how Debye shielding and Landau damping were derived from equations (\ref{phihatL})-(\ref{eps}) in reference \cite{DbCb}. These equations are obtained by introducing the continuous velocity distribution after particle dynamics is taken into account, and not before, as occurs when kinetic equations are used. However, they look very similar to those including initial conditions in Landau contour calculations of Langmuir wave growth or damping, usually obtained by linearizing Vlasov equation and using Fourier-Laplace transform, as described in many textbooks. Indeed, the smoothed self-consistent potential $\varphi_{\rm cg}$
is determined by the response function $\epsilon(\bfm,\omega)$, viz.\  the classical plasma dielectric function, whose Vlasovian expression involving $\partial f_0 / \partial \bfv$ obtains by a mere integration by parts if $f_0$ is differentiable.
For a cold plasma, $\epsilon(\bfm,\omega) = 1 - {\omega_{\rmpp}^2}/{\omega^2}$, as expected (however this case is ruled out since the present theory is correct only for a large number of particles in the Debye sphere, as recorded in the previous section).

In equation (\ref{phihatL}), $\varphi_{\rm lin}^{(\rm{bal})}(\bfm,\omega)$ is a discrete sum over the $N$ particles. As computed in \cite{Gasio,Bal,Rost}, in the inverse Laplace transform of $\varphi_{\rm cg}(\bfm,\omega)$, the pole $\omega = \bfk_{\bfm}  \cdot \dot \bfr_j(0)$ of $\varphi_{\rm lin}^{(\rm{bal})}(\bfm,\omega)$ brings a contribution to $\varphi_{\rm cg}(\bfm,t)$ which is the shielded Coulomb potential of particle $j$. This potential is often computed in textbooks by adding a test particle to a Vlasovian plasma (see for instance chapter 9 of \cite{Nicholson}). Since Debye shielding is effective over distances larger than the Debye length, this discussion is relevant under the assumption that our box size is large in comparison,
viz.\ $L \gg \lambda_\rmD$.

As in the Vlasovian approach, the part of $\varphi_{\rm cg}(\bfm,t)$ due to dominant poles of $1/\epsilon(\bfm,\omega)$ corresponds to one Langmuir wave propagating in the direction of $\bfk_{\bfm}$ and one propagating in the opposite direction,
with the corresponding Landau growth or damping.
Therefore, \emph{the full potential in the plasma turns out to be the sum of two contributions only~: one due to Langmuir waves and one due to all the shielded Coulomb potentials of individual particles.}

Equation (\ref{phihatL}) becomes the Vlasovian one on replacing in equation\ (\ref{*phij0hat}) the discrete sums over particles with integrals over the smooth distribution function $f(\textbf{r},\textbf{v})$. This yields
\begin{equation}
  \varphi_{\rm cg}^{(\rm{bal})}(\bfm,\omega)
  = - \frac{\rmi e}{\epsilon_0 k_{\bfm}^2} \int
    \frac{f_1(\bfm,\bfv)}
         {\omega -\bfk_{\bfm} \cdot \bfv} \
     \rmd^3 \bfv,
\label{phi0hatcg}
\end{equation}
the Vlasovian formula for the ballistic potential. The $N$-body description reveals that the contribution of $\varphi_{\rm cg}^{(\rm{bal})}(\bfm,\omega)$ to $\varphi_{\rm cg}(\bfm,t)$ in the Vlasovian approach is the continuous limit of the sum of all the Debye-shielded potential of the particles. In this approach, this contribution is only a transient \cite{HW}. Though its physical meaning is unclear, \emph{it does not correspond to a robust supplementary contribution to the full electrostatic potential in the plasma which would add to the potential of Langmuir waves and to the shielded Coulomb potentials of particles.}

To afford the description of nonlinear effects in wave-particle dynamics, section 6.1 of reference \cite{DbCb} generalizes the linear analysis analogous to that in section \ref{FEP} of the present paper by applying it to the particles which are not resonant with Langmuir waves, while keeping the exact contribution to the electrostatic potential of the remaining $N'$ tail particles that may experience trapping or chaotic dynamics. This
eventually leads to equations describing the self-consistent dynamics of $M$ Langmuir waves and of $N'$ tail particles (these equations could be also obtained with the formalism of the present paper). For a one-dimensional plasma, the linear theory of Langmuir waves can be formulated on this basis, with the bonus that the information on particle dynamics is provided in parallel with the waves' \cite{EZE,EEbook,EFields,Houches}. This approach also provides spontaneous emission \cite{EZE,EEbook,DbCb}, and some nonlinear effects like damping with trapping \cite{FirEl2}, and the decoupling of the waves from the strongly chaotic motion of the particles when there is a plateau in the particle velocity distribution function (see sec.\ 2.2 of \cite{BEEB}).

\section{Comparison with Pines and Bohm 1952}
\label{BGP}

In 1952, Pines and Bohm showed that an $N$-body approach can bring a lot of insight into basic plasma effects \cite{PiBo}. We now summarize the part of their results overlapping with our theory. Pines and Bohm already considered a One Component Plasma in a periodic box and wrote the equations of motion of its $N$ electrons acted upon by their electrostatic potential. Working with the physical time $t$, they got an equation (number (8) in their paper) for the time evolution of a Fourier component $\rho_{\bfk}$ of the charge density. This equation involves a contribution of all Fourier components whose structure is similar to equation (\ref{*phihat}), but \emph{without linearizing particle dynamics}. In order to neglect the contribution of all other Fourier components, they introduced the now celebrated ``random phase approximation" by showing that the particle positions may be considered as nearly random for all times if there are many particles in a Debye sphere of a plasma in thermal equilibrium.

As a first step, they considered the collective part of $\rho_{\bfk}$, and showed that,
for small wavenumbers $\| \bfk \|$, the charge density oscillates at the plasma frequency,
and that plasma oscillations exist for $\|\bfk\| \ll \lambda_\rmD^{-1}$. Inductively, they derived a harmonic equation for non-vanishing $\|\bfk\|$'s with an oscillation frequency $\omega$ assumed implicitly real, and satisfying a discrete version of the Vlasovian dispersion relation (their equation (23)). They recovered the Bohm-Gross dispersion relation for small $\|\bfk\|$'s.

As a second step, in their section III they split $\rho_{\bfk}$ into a collective and a granular part. The evaluation of the latter for thermal particles yielded correctly the Yukawa version of Debye shielding.

In contrast, as those presented in most textbooks, our approach to Debye shielding and Landau damping is a \emph{linear theory}. Laplace transform can then be applied readily. As already demonstrated by Landau's theory, the Laplace transform is a powerful heuristic tool. In the $N$-body approach, it shows the full potential in the plasma to split naturally in two contributions related to two families of poles~: one due to Langmuir waves subjected to Landau growth and damping, and one due to all the shielded Coulomb potentials of individual particles, whatever be their velocity. By considering the initial $N$-body distribution as close to a set of monokinetic arrays of particles, it was shown in Appendix B of reference \cite{DbCb} that linearization provides naturally the cancellation of non-diagonal terms in equation (\ref{*phihat})~; \emph{it is not necessary to use the random phase approximation}, or other statistical arguments.

Particle deflection and Gauss' theorem were used by Pines and Bohm to interpret their Yukawa version of Debye shielding. Section 5 of reference \cite{DbCb} brings more mechanical insight to this interpretation. There, instead of applying the Laplace transform to linearized particle dynamics, one applies Picard iteration technique (a standard method to prove the existence and uniqueness of solutions to first-order equations with given initial conditions) to the equation of motion of a particle, say $P$, due to the Coulomb forces of all other ones \emph{without linearizing particle dynamics}. This calculation shows that particle $P$ acts on another particle $P'$, directly by its Coulombian force, and indirectly by deflecting the orbits of all other particles whose force on particle $P'$ is modified by this deflection. The direct Coulombian force is reduced by the indirect action, since the Coulombian deflections due to particle $P$ diminish the number of other particles in a sphere surrounding it, which diminishes the apparent charge of particle $P$ according to Gauss' theorem. This shielding effect becomes dominant at a distance on the order of $\lambda_\rmD$.

When starting from random particle positions, the typical time-scale for shielding to set in is the time for a thermal particle to cross a Debye sphere, i.e.\ $\omega_{\rmpp}^{-1}$.
This is also the time for the equilibrium pair-correlation function to settle at its time-asymptotic values.
Therefore, \emph{shielding results from the accumulation of almost independent repulsive deflections with the same qualitative impact on the effective electric field of particle $P$.} If point-like ions were present, the attractive deflection of charges with opposite signs would have the same effect. \emph{The Coulombian deflections induce a self-organization of the plasma~: it becomes a dielectric.} So, these deflections do not produce only the disorder leading to collisional transport, but also the order corresponding to shielding and to the behaviour as a dielectric.
Paradoxically, ``collisionless'' Landau damping turns out to occur because of what is usually called ``collisions". Since the global deflection of particles includes the contributions of many other ones, the density of the electrons does not change, at variance with what occurs for the shielding at work next to a Langmuir probe. Gauss' theorem was used in section 2.2.1 of \cite{APiel} to provide an intuitive explanation of the shielding of such a probe.

\section{Conclusion}
\label{Concl}

This paper provided further insight into the unified $N$-body description of Debye shielding and Landau damping. First, by deriving the fundamental equation for the electrostatic potential of reference \cite{DbCb} in a simpler and more rigorous way. Second, by stressing that the term from initial conditions in Landau's calculation of Langmuir waves does not bring new physics, but is a mere damped continuous version of the Debye shielded potentials of all particles whose physical meaning is unclear in the Vlasov picture. Third, by discussing more precisely the condition for approximating the $N$-body distribution by a smooth one in the dispersion relation. Fourth, by comparing the present approach with the 1952 one by Pines and Bohm \cite{PiBo} which already brought a lot of insight into Debye shielding and Langmuir waves. In particular, we have shown how the Coulombian deflections of particles bring order to the plasma by producing Debye shielding and by making the plasma behave as a dielectric. Furthermore, the Laplace transform confirms to be a powerful heuristic tool providing a compact and straightforward simultaneous derivation of Debye shielding and Landau damping. Linearization was shown to provide naturally the cancellation of non-diagonal terms in the fundamental equation for the electrostatic potential. This avoids using the random phase approximation or statistical arguments.

The $N$-body approach provides a derivation of Debye shielding for a single mechanical realization of the plasma. This was used in an $N$-body calculation of collisional transport of the One Component Plasma (OCP) incorporating all impact parameters with no ad hoc cut-off \cite{EED14}. This calculation explains why a two-body calculation yields a correct estimate of collisional transport, while most of this transport is due to the simultaneous action of many particles with impact parameters between the inter-particle distance and the Debye length.

When generalizing this theory to the collisional transport of a plasma with granular electrons and ions, the singularity of head-on collisions of particles with opposite charges can be avoided by excluding the case of a vanishing impact parameter which has probability 0. Then, the deflection of particles with same sign is just the opposite of the deflection of particles with opposite signs, so that the classical Rutherford cross-section is the same for charges of like sign and opposite signs. However, one might wonder whether a quantum mechanical treatment would not modify the scattering cross-section. Yet, it does not. This result, traditionally derived with the Born approximation, was obtained more recently with a much simpler calculation using Fermi golden rule that makes intuitive why the classical and quantum results agree \cite{Gauth}.

The theory of collisional transport of reference \cite{EED14} and the approach of the present paper show that $N$-body dynamics, which has always been the ultimate reference in plasma textbooks, is now also a practical tool.
Furthermore, Laplace's dream was not a mere utopia, since classical mechanics can genuinely describe non trivial aspects of the macroscopic dynamics of a many-body system.

One of us (DFE) thanks the Program Committee of the 42st EPS Conference on Plasma Physics and G.L.~Delzanno to have led him to look for a more pedagogical derivation of the results of reference \cite{DbCb} in order to prepare an invited talk and a course~; this was the origin of section \ref{FEP} of this paper.
He also thanks J.~Daligault for recording him the pioneering 1952 contribution of Pines and Bohm \cite{PiBo}
to the $N$-body approach to basic plasma physics,
and J.~Daligault and S.~Khrapak for a critical reading of our manuscript. 
We thank the two anonymous referees for leading us to formulate our theory in a more rigorous way, and to write a clearer and more self-contained paper.

\appendix
\section{Full derivation}
\label{CDer}

We now consider $\varphi(\bfr)$, the potential created by the $N$ particles at any point where there is no particle. Its discrete Fourier transform is readily obtained from the Poisson equation, and is given by $\tilde{\varphi}(\mathbf{0}) = 0$, and for $\bfm \neq \mathbf{0}$ by
\begin{equation}
  \tilde{\varphi}(\bfm)
  = -\frac{e}{\epsilon_0 k_{\bfm}^2} \sum_{j = 1}^N
     \exp[- \rmi \bfk_{\bfm} \cdot \bfr_j(t)],
\label{phitildetotM}
\end{equation}
where $\bfr_j(t)$ is the position at time $t$ of particle $j$ acting as a source,
$\tilde{\varphi}(\bfm)
= \int \varphi(\bfr) \exp(- \rmi \bfk_{\bfm} \cdot \bfr) \, \rmd^3 \bfr$,
with 
$\bfk_{\bfm} = \frac{2 \pi}{L} \, \bfm$, and $k_{\bfm} = \|\bfk_{\bfm}\|$.

Let $\bfr_{j0}$ and $\bfv_{j}$ be respectively the initial position and velocity of particle $j$,
and let $\Delta \bm{r}_j (t)= \bfr_j (t) - \bfr_{j0} - \bfv_j t$. We now compute a perturbative solution to the full $N$-body dynamics and the resulting Fourier components of the potential for
 $\bfm$ running in the finite domain $\cMs$ defined in subsection \ref{CEEP},
such that $k_{\bfm} b_{\mathrm{smooth}} \leq 1$.

To this end, setting $\bfr_j = \bfr_{j0} + \bfv_j t + \Delta \bm{r}_j (t)$ in equation (\ref{phitildetotM}),
we replace $\tilde{\varphi}$ with its expansion to first order in the $\Delta \bm{r}_j (t)$'s
\begin{equation}
\tilde{\varphi}_{\rm lin} (\bfm,t)
  =
  - \sum_{j = 1}^N
  \frac{ e}{\epsilon_0 k_{\bfm}^2} \exp [- \rmi \bfk_{\bfm} \cdot (\bfr_{j0} + \bfv_{j} t)] \ [1 - \rmi \bfk_{\bfm} \cdot \Delta \bm{r}_j(t)],
\label{phitildnj}
\end{equation}
whose time Laplace transform is
\begin{equation}
\varphi_{\rm lin} (\bfm,\omega)
  =
  - \sum_{j = 1}^N
  \frac{ e}{\epsilon_0 k_{\bfm}^2} \exp [- \rmi \bfk_{\bfm} \cdot \bfr_{j0} ] \
  [ \frac{\rmi}
             {\omega -\bfk_{\bfm}  \cdot \bfv_j} - \rmi \bfk_{\bfm} \cdot \Delta \bm{r}_j (\omega - \bfk_{\bfm}  \cdot \bfv_j) ],
\label{Lphitildnj}
\end{equation}
where the Doppler shift $- \bfk_{\bfm} \cdot \bfv_j$ comes from the linear dependence on $t$ of the exponent of equation\ (\ref{phitildnj}).

To compute $\Delta \bm{r}_j (\omega - \bfk_{\bfm}  \cdot \bfv_j) $, we use Newton's equation for the particles 
\begin{equation}
  \ddot{\bfr}_j
  = \frac{e}{m_\rme} \nabla \varphi_{\rm lin} (\bfr_j),
\label{rsectot}
\end{equation}
where $m_\rme$ is the electron mass.
When $\bfm$ runs over the finite domain such that $k_{\bfm} b_{\mathrm{smooth}} \leq 1$,
the self-field due to $\varphi$ vanishes, and it is not necessary to exclude self-interactions.
Therefore, one may use the harmonics of the electrostatic potential due to all particles defined by equation\ (\ref{phitildetotM}),
which yields
\begin{equation}
\varphi_{\rm lin}(\bfr_j) = \frac{1}{L^3}\sum_{\bfm, \, k_{\bfm} b_{\mathrm{smooth}} \leq 1} \tilde{\varphi}_{\rm lin} (\bfm) \exp(\rmi \bfk_{\bfm} \cdot \bfr_j) ,
\label{phiInv}
\end{equation}
where $\mathbf{r}_j$ now is the position of particle $j$ as a massive object subjected to a force. Using equation\ (\ref{phitildnj}), the linearized particles dynamics defined by equation\ (\ref{rsectot}) is then given by
\begin{equation}
  \Delta \ddot{\bm{r}}_j
  =
  \frac{\rmi e}{L^3 m_\rme} \sum_{\bfn, \, k_{\bfn} b_{\mathrm{smooth}} \leq 1} \bfk_{\bfn} \
    \tilde{\varphi}_{\rm lin}(\bfn,t) \exp[\rmi \bfk_{\bfn} \cdot (\bfr_{j0} + \bfv_{j} t)] ,
\label{delrsec}
\end{equation}
Since $\Delta \bm{r}_j =0$ and $\Delta \dot{\bm{r}}_j = 0$ at $t = 0$, the time Laplace transform of equation\ (\ref{delrsec}) is
\begin{equation}
  - \omega^2 \Delta \bm{r}_j(\omega)
  = \frac{\rmi e}{L^3 m_\rme} \sum_{\bfn, \, k_{\bfn} b_{\mathrm{smooth}} \leq 1}
                \bfk_{\bfn} \exp(\rmi \bfk_{\bfn} \cdot \bfr_{j0})
                  \ \varphi_{\rm lin}(\bfn,\omega + \bfk_{\bfn} \cdot \bfv_{j}),
\label{rLapl}
\end{equation}
where the Doppler shift $\bfk_{\bfn} \cdot \bfv_{j}$
comes from the linear dependence on $t$ in the exponent of equation (\ref{delrsec}).
Computing $\Delta \bm{r} (\omega - \bfk  \cdot \bfv) $ in equation (\ref{Lphitildnj}) from the expression of $\Delta \bm{r}(\omega)$ given by equation (\ref{rLapl}) yields the fundamental equation (\ref{*phihat}) for the potential.
Note that the assumption $k_{\bfm} b_{\mathrm{smooth}} \leq 1$,
with $b_{\mathrm{smooth}} \ll \lambda_\rmD$, defining the set $\cMs$,
\emph{excludes scales which are irrelevant to Debye shielding and Landau damping,
since these phenomena involve scales larger than, or of the order of $\lambda_\rmD$}.



\end{document}